\documentclass[twocolumn,showpacs,preprintnumbers,pra,amsmath,amssymb]{revtex4}

\usepackage{graphicx}
\usepackage{bm}
\begin{document}
\title{No-signaling Quantum Key Distribution: Solution by Linear Programming}

\author{Won-Young Hwang$^{1,2}$ \footnote{Email: wyhwang@jnu.ac.kr}, Joonwoo Bae$^{3}$, and Nathan Killoran$^{2}$}

\affiliation{$^{1}$Department of Physics Education, Chonnam National University, Gwangju 500-757, Republic of Korea\\
$^{2}$ Institute for Quantum Computing and Department of Physics \& Astronomy, University of Waterloo, Waterloo N2L 3G1, Canada\\
$^{3}$ Centre for Quantum Technologies, National University of Singapore, Singapore 117542, Singapore\\
}
\begin{abstract}
We outline a straightforward approach for obtaining a secret key rate using only no-signaling constraints and linear programming. Assuming an individual attack, we consider all possible joint probabilities. Initially, we study only the case where Eve has binary outcomes, and we impose constraints due to the no-signaling principle and given measurement outcomes. Within the remaining space of joint probabilities, by using linear programming, we get bound on the probability of Eve correctly guessing Bob's bit. We then make use of an inequality that relates this guessing probability to the mutual information between Bob and a more general Eve, who is not binary-restricted. Putting our computed bound together with the Csisz\'ar-K\"orner formula, we obtain a positive key generation rate. The optimal value of this rate agrees with known results, but was calculated in a more straightforward way, offering the potential of generalization to different scenarios.
\end{abstract}
\maketitle
\section{Introduction}
A nonlocal realistic model, the de Broglie-Bohm theory, is not only consistent with quantum theory but also coherently describes measurement processes including wave-function collapse \cite{Boh93}. This raises a question if all realistic models must be nonlocal to be consistent with quantum theory, which led to the discovery of Bell's inequality \cite{Bel64, Bel87}.

Recently, the nonlocality involved with Bell's inequality and entanglement has entered a new phase of its development. It turned out that entanglement is a concrete physical resource for information processing \cite{Nie00}. In the same context, interestingly, it was found that with nonlocal correlations we can generate a cryptographic key, a private random shared sequence, whose security relies on only the no-signaling principle \cite{Bar05,Aci06,Aci06-2}. For this, no quantum theory is used for the security analysis. However, the only currently available way to realize nonlocal correlations is by using quantum entanglement. So these protocols are called no-signaling {\it quantum} key distribution (QKD). Remarkably, what is used to show security in no-signaling QKD is only the outcomes of measurements. As long as the outcomes satisfy a certain condition, security is provided, no matter how the outcomes are generated.
Thus, no-signaling QKD has device-independent security. To satisfy the security condition, detector efficiency must be much higher than what is currently achievable.


In Refs. \cite{Aci06,Aci06-2}, the security of no-signaling QKD against individual attacks has been analyzed in a novel way, exploiting the intrinsic structure of no-signaling probabilities \cite{Bar05-2,Jon05}. In particular, by fixing the size of the input and output alphabets that are used to generate a secret key between the legitimate parties, a finite set of extremal points are distinguished. Information about the no-signaling polytope structure leads to huge simplifications in the security analysis. In Ref. \cite{Paw12}, security against individual attacks was shown using the insight that no-signaling and non-local probabilities are generally monogamous. Indeed, a monogamy relation that is valid for no-signaling probabilities is explicitly employed to show the security. This approach can be applied even if the eavesdropper's alphabet is not binary \cite{Hwa12, Paw12}.

In this paper, we present a security analysis of no-signaling QKD protocols by numerically optimizing no-signaling probabilities. This explicitly shows that direct optimization over no-signaling probabilities can be used as the main theoretical tool to prove security. Specifically, we consider the protocol proposed by Acin, Massar, Pironio (AMP) \cite{Aci06-2, AMP}. To motivate the advantage of our approach, we note that the method in Refs. \cite{Aci06,Aci06-2} relies on the specific structure of certain no-signaling polytopes shown in Refs. \cite{Bar05-2,Jon05}. However, it seems that the generalization to larger alphabets or higher dimensions is much harder to analyze; see for instance Ref. \cite{Pir11}. Nevertheless, our result provides a straightforward enhancement to the analysis of no-signaling probabilities, and the formalism could potentially be applied to even more complicated scenarios.

This paper is organized as follows. First we consider Eve's (an eavesdropper's) guessing probability about Bob's (a receiver's) bit. That is, we consider the case that Eve's outcomes are binary. Within the remaining space of joint probabilities, we maximize $P_{E}$, the probability that Eve correctly guesses Bob's bit, by linear programming. Then, we derive a bound on the mutual information between Bob and a general Eve (whose number of outcomes is now unrestricted), $I_{BE}$, by using the maximal $P_{E}$. A key generation rate $K$ is obtained by using the Csisz\'ar-K\"orner formula \cite{Csi78}. In our case, $K= I_{AB}-I_{BE}$, where $I_{AB}$ is the mutual information between Alice (a sender) and Bob.
\section{Main contents}
\subsection{AMP protocol}
Two users, Alice and Bob, attempt to distribute a Bell state, $|\phi^+\rangle= (1/\sqrt{2})(|0\rangle_A |0\rangle_B+ |1\rangle_A|1\rangle_B) $, where $A$ and $B$ denote Alice and Bob, respectively, and $|0\rangle$ and $|1\rangle$ compose an orthonormal basis of a quantum bit (qubit). To mimic a realistic case with channel noise, we assume the Bell state was transformed to a Werner state
\begin{equation}
\rho= p |\phi^+\rangle \langle \phi^+| +(1-p) \frac{I}{4},
\label{1}
\end{equation}
where $0 \leq p \leq 1$. Although we use the Werner state to model potential data, our method does not rely on this. For each copy of the distributed state, Alice chooses the value of an index $x$ among $0, 1$, and $2$  with probabilities $q$, $(1-q)/2$, and $(1-q)/2$, respectively. Then she performs a measurement $M_x$ on her qubit. $M_0$ is a measurement composed of the projections $\{|+\rangle \langle +|, |-\rangle \langle -|\}$ where $|\pm \rangle= (1/\sqrt{2})(|0\rangle \pm |1\rangle) $. $M_1$ and $M_2$ are measurements composed of $\{|\pi/4 \rangle \langle \pi/4|, |5\pi/4\rangle \langle 5\pi/4|\}$ and $\{|-\pi/4 \rangle \langle -\pi/4|, |-5\pi/4\rangle \langle -5\pi/4|\}$, respectively. Here, $|\phi \rangle= (1/\sqrt{2})(|0\rangle+ e^{i\phi} |1\rangle)$ is a state obtained by rotating the state $|+\rangle $ around the $z$-axis by an angle $\phi$. Bob also chooses a value of his index $y$ for each copy, either $0$ or $1$, with probabilities $q'$ and $1-q'$, respectively. Then he performs a measurement $N_y$ on his qubit. Here $N_0=M_0$ and $N_1$ is composed of $\{|\pi/2 \rangle \langle \pi/2|, |3\pi/2\rangle \langle 3\pi/2|\}$. Next, both Alice and Bob publicly announce their values $x$ and $y$ for each copy. Measurement outcomes in the case $x=y=0$ are kept and used to generate the key. Outcomes from other cases are publicly announced to estimate Eve's information. Alice and Bob choose $q$ and $q'$ close to $1$ so that almost events are in the case $x=y=0$. This does not affect the security in the asymptotic case we consider.
\subsection{Constraints on the probability distributions}
We assume an individual attack in which Eve follows the same procedure for each instance. For each choice of measurements $x$ and $y$ by Alice and Bob, there is a joint probability for measurement outcomes $a,b,e$ for Alice, Bob, and Eve, respectively. The joint probability for $a,b,e$, conditioned on measurements $x$ and $y$ is denoted by $P(a,b,e|x,y)$.
Here, $a$ and $b$ are binary variables according to the protocol. The number of Eve's outcomes $e$ should be arbitrary in principle. However, for now we consider the case that Eve's outcome is binary. We do this because we are interested in the guessing probability, and Eve's final guess has to be binary to match Bob's alphabet.

Let us write constraints for the joint probabilities. First, they satisfy normalization
\begin{equation}
\sum_{a,b,e} P(a,b,e|x,y)= 1
\label{2}
\end{equation}
for each $x,y$. Let us denote the marginal distribution for Alice and Bob, $\sum_{e} P(a,b,e|x,y)$, by $P(a,b,\triangle|x,y)$.

The marginal distributions corresponding to the state in Eq. (\ref{1}) should be consistent with the measurement outcomes. For the measurement basis choice $(x=0,y=0)$  we have
\begin{eqnarray}
 P(0,0,\triangle|0,0) &=& P(1,1,\triangle|0,0)= \frac{p}{2}+ \frac{1-p}{4},
 \nonumber\\
 P (0,1,\triangle|0,0) &=& P(1,0,\triangle|0,0)=  \frac{1-p}{4}.
\label{3}
\end{eqnarray}
For $(x=0,y=1)$, where there is no correlation,
\begin{eqnarray}
P(a,b,\triangle|0,1)= \frac{1}{4}
\label{4}
\end{eqnarray}
for each $a$ and $b$. For $(x=1,y=0)$, $(x=1,y=1)$, and  $(x=2,y=0)$,
\begin{eqnarray}
 P(0,0,\triangle|x,y) &=& P(1,1,\triangle|x,y) \nonumber\\
  &=& 0.854 \hspace{2mm} \frac{p}{2} + \frac{1-p}{4} \equiv \alpha
 \nonumber\\
 P (0,1,\triangle|x,y) &=& P(1,0,\triangle|x,y) \nonumber\\ &=& 0.146 \hspace{2mm} \frac{p}{2}+ \frac{1-p}{4}\equiv \beta,
\label{5}
\end{eqnarray}
where the two numerical values, $0.854$ and $0.146$, are obtained from measurement outcomes for the Bell state.
For $(x=2,y=1)$,
\begin{eqnarray}
 P(0,0,\triangle|2,1) =P(1,1,\triangle|2,1) &=& \beta
 \nonumber\\
 P (0,1,\triangle|2,1)=P(1,0,\triangle|2,1)&=& \alpha.
\label{6}
\end{eqnarray}

Now we consider no-signaling conditions. Because the marginal distribution for Alice and Eve must be independent of Bob's basis choice, we have
\begin{equation}
 P(a,\triangle,e|x,0)= P(a,\triangle,e|x,1)
 \label{7}
\end{equation}
for each $x$. Here we use a notation for marginal distributions analogous to the previous one. Similarly,
\begin{equation}
 P(\triangle,b,e|0,y)= P(\triangle,b,e|1,y) =P(\triangle,b,e|2,y)
 \label{8}
\end{equation}
for each $y$. Another no-signaling constraint is that Eve's marginal distribution is independent of the basis choices of Alice and Bob,
\begin{equation}
 P(\triangle,\triangle,e|x,y)= P(\triangle,\triangle,e|0,0)
 \label{9}
\end{equation}
for each $x,y$.
\subsection{Maximizing guessing probability $P_{E}$}
Here we maximize the guessing probability, $P_{E}$, for a binary-restricted Eve within these constraints (\ref{2})-(\ref{9}) by linear programming.

For visual convenience, $P(a,b,e|x,y)$ are denoted as:
\begin{eqnarray}
P(a,b,e|0,0) &=& x_{abe}, \hspace{3mm} P(a,b,e|0,1)= y_{abe}, \nonumber\\
P(a,b,e|1,0) &=& z_{abe}, \hspace{3mm} P(a,b,e|1,1)= u_{abe}, \nonumber\\
P(a,b,e|2,0) &=& v_{abe}, \hspace{3mm} P(a,b,e|2,1)= w_{abe}.
\label{A1}
\end{eqnarray}
We regard $abe$ as a binary number, for example, $P(1,0,1|0,0)= x_{101}= x_5$.

Now let us rewrite the constraints regarding measurement outcomes. For Eqs. (\ref{3}) and (\ref{4}), we have, respectively,
\begin{eqnarray}
 x_0+ x_1 &=& x_6 +x_7= \frac{p}{2}+ \frac{1-p}{4},
 \nonumber\\
 x_2+ x_3 &=& x_4 +x_5= \frac{1-p}{4},
\label{A2}
\end{eqnarray}
and
\begin{eqnarray}
 y_0+ y_1= y_2 +y_3= y_4+ y_5= y_6 +y_7= \frac{1}{4}.
\label{A3}
\end{eqnarray}
For Eq. (\ref{5}), we have
\begin{eqnarray}
A_0+ A_1 &=& A_6 +A_7=\alpha \nonumber\\
A_2+ A_3 &=& A_4 +A_5=\beta,
\label{A4}
\end{eqnarray}
where $A=z,u,v$. For Eqs. (\ref{6}), we have
\begin{eqnarray}
w_0+ w_1 &=& w_6 +w_7=\beta \nonumber\\
w_2+ w_3 &=& w_4 +w_5=\alpha.
\label{A5}
\end{eqnarray}
We can see that Eqs. (\ref{A2})-(\ref{A5}) make the normalization in Eq. (\ref{2}) satisfied. Thus the normalization condition is redundant and can be removed.

The no-signaling condition in Eq. (\ref{7}) can be expressed as
\begin{eqnarray}
x_{i}+ x_{i+2} &=& y_{i}+ y_{i+2}, \nonumber\\
z_{i}+ z_{i+2} &=& u_{i}+ u_{i+2}, \nonumber\\
v_{i}+ v_{i+2} &=& w_{i}+ w_{i+2},
\label{A6}
\end{eqnarray}
where $i=0,1$ and $4,5$. We can see that, by Eqs. (\ref{A2})-(\ref{A5}), the case when $i=0,4$ implies the case when $1,5$, respectively. Thus the latter cases can be removed.
The no-signaling condition in Eq. (\ref{8}) can be expressed as
\begin{eqnarray}
x_{j}+ x_{j+4} &=& z_{j}+ y_{j+4}, \nonumber\\
z_{j}+ z_{j+4} &=& v_{j}+ v_{j+4}, \nonumber\\
y_{j}+ y_{j+4} &=& u_{j}+ u_{j+4}, \nonumber\\
u_{j}+ u_{j+4} &=& w_{j}+ w_{j+4},
\label{A7}
\end{eqnarray}
where $j=0,1,2,3$. We can also see that, by Eqs. (\ref{A2})-(\ref{A5}), the case when $j=0,2$ implies the case when $j=1,3$, respectively. Thus the latter cases are redundant and can be removed. We can verify that Eqs. ({A6}) and ({A7}) (or equivalently, Eqs. ({7}) and ({8})) lead to Eq. (\ref{9}), which can thus be removed. As a result, we can remove all variables $B_i$ where $B=x,y,z,u,v,w$ and $i$ is an odd number.

Therefore, by non-negativity of each quantity, the space in which we optimize $I_{BE}(2)$ is as follows:
\begin{eqnarray}
0 \leq x_{k} \leq \frac{p}{2}+ \frac{1-p}{4}, \hspace{5mm}
0 \leq x_{l} \leq \frac{1-p}{4},
\label{A8}
\end{eqnarray}
where $k=0,6$ and $l=2,4$,
\begin{eqnarray}
0 \leq y_{j} \leq \frac{1}{4},
\label{A9}
\end{eqnarray}
where j=0,2,4,6, and
\begin{eqnarray}
0 \leq A_{k} \leq \alpha, &&
0 \leq A_{l} \leq \beta, \nonumber\\
0 \leq w_{k} \leq \beta,  &&
0 \leq w_{l} \leq \alpha,
\label{A10}
\end{eqnarray}
where $A=z,u,v$ and $k=0,6$ and $l=2,4$. The constraints are those that remain in Eqs. (\ref{A6}) and (\ref{A7}) after removing odd numbered variables.

Because the key is generated only from the results where $x=y=0$, we need to consider the joint distribution $P(\Delta,b,e|0,0) \equiv R(b,e)$. Now, the guessing probability
\begin{eqnarray}
 P_E&=& R(0,0)+ R(1,1)= x_0+ x_4+ x_3+x_7 \nonumber\\
    &=& (x_0+ x_4)- (x_2+ x_6)+ \frac{1}{2},
 \label{A10-2}
\end{eqnarray}
where Eqs. (\ref{A1}) and (\ref{A2}) are used.

To maximize the guessing probability, we use linear programming \cite{Gas10}. First we note that the constraints (\ref{A8})-(\ref{A10}) define a convex set. We define $\mathcal{C}\in[0,\tfrac{1}{2}]\times[0,\tfrac{1}{2}]$ as the projection of this set onto the $(a,b)$-plane where $a\equiv x_0+ x_4$ and $b\equiv x_2+ x_6$. We notice that $\mathcal{C}$ is convex and is symmetric under transformations $(a,b)\leftrightarrow (b,a)$ and $(a,b)\leftrightarrow (\tfrac{1}{2}-a,\tfrac{1}{2}-b)$. In our case the linear function $P_E$ can be directly optimized using linear programming. Specifically, for fixed value of the noise parameter $p$, we perform the following optimization:
\begin{eqnarray}
P_E^\text{max}=
\begin{cases}
	\max &~P_E(a,b)\\
	\text{subject to } & (a,b)\in \mathcal{C}.
\end{cases}
\end{eqnarray}
\subsection{Maximizing $I_{BE}$}
Now we obtain a bound on the mutual information, $I_{BE}$, from the guessing probability $P_{E}$. There is a simple relation for the problem \cite{Paw12}: Let us consider a marginal distribution for Bob and Eve, $R(i,j)$. Here Eve is not binary-restricted ($i=0,1$ and $j=0,1,2,...$). Consider conditional probabilities $P(0|j)$ and $P(1|j)$ due to the joint probability $R(i,j)$. The joint probabilities can be written as $R(i,j)= P(i|j) P(j)$, where $P(j)$ is a marginal distribution for Bob. The mutual information is
\begin{eqnarray}
I_{BE} &=& H(i)- H(i|j)
 \nonumber\\
      &=& H(i)- \sum_{j} H[P(0|j)] P(j),
\label{B1}
\end{eqnarray}
where the binary entropy function $H[q] \equiv -[q \log_2 q+ (1-q) \log_2 (1-q)]$ has been introduced. Let $P_E(j)$ be Eve's probability to guess Bob's outcome correctly, when her outcome is $j$. However, we can observe that $P_E(j)= \max\{P(0|j), P(1|j)\}$. Because $H[P(0|j)]= H[P(1|j)]=H[P_E(j)]$ here, we have
\begin{eqnarray}
I_{BE} &=& H(i)- \sum_{j} H[P_E(j)] P(j),
\label{B2}
\end{eqnarray}
The (average) guessing probability is $P_E= \sum_{j} P_E(j) P(j)$. However, for a fixed $P_E$, the smallest value of the quantity $\sum_{j} H[P_E(j)] P(j)$ is obtained when each $P_E(j)$ take either $1/2$ or $1$, by the concavity of the binary entropy as discussed in Ref. \cite{Paw12}. Let $r$ denote the sum of all $P(j)$ such that $P_E(j)= 1/2$. Then we have $P_E= 1-(r/2)$ and thus
\begin{eqnarray}
\sum_{j} H[P_E(j)] P(j) \geq r = 2(1-P_E).
\label{B3}
\end{eqnarray}
Now we obtain
\begin{eqnarray}
I_{BE} &=& H(i)- \sum_{j} H[P_E(j)] P(j) \nonumber\\
&\leq& 1- 2(1-P_E)= 2P_E-1,
\label{B4}
\end{eqnarray}
where the constraint $H(i)=1$ is used. Therefore we get
\begin{eqnarray}
I_{BE} \leq 2 P_E-1.
\label{B5}
\end{eqnarray}
\begin{figure}
\includegraphics[width=8cm]{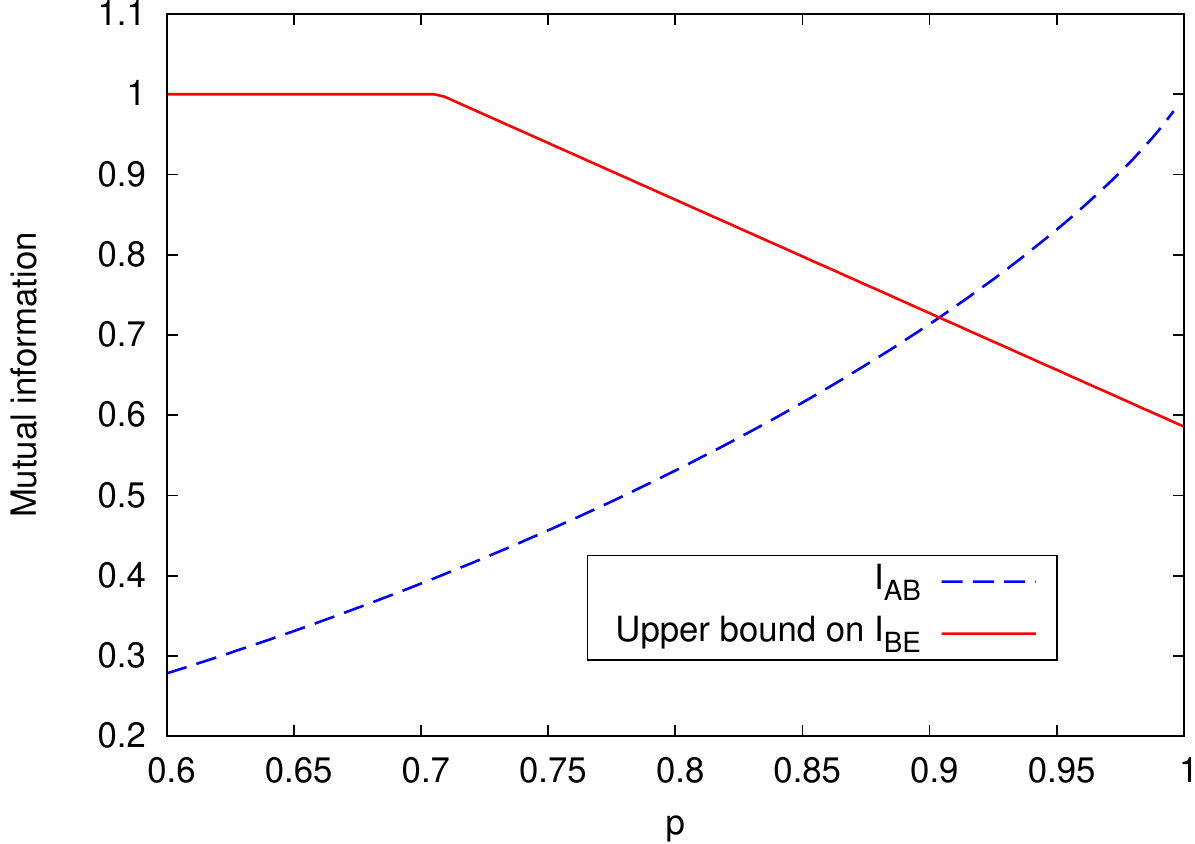}
\caption{Mutual informations depending on the noise parameter $p$. Positive key is possible in the region where $I_{BE}$ is smaller than $I_{AB}$ (dashed line). 
}
\label{Fig-1}
\end{figure}

Using the relation (\ref{B5}) and the maximal guessing probability obtained by linear programming, we can get a bound on $I_{BE}$ as shown in Fig. 1.  Then, by the Csisz\'ar-K\"orner formula \cite{Csi78}, we can get a lower bound on the key generation rate $K=  I_{AB}-I_{BE}$. As we can see, in the regime $p< \tfrac{1}{\sqrt{2}}$ where the Werner state admits a local realistic model, Eve has full information about Bob, namely $P_{E}=1$, so there can be no secret key. However, in the regime where $\tfrac{1}{\sqrt{2}} \leq p \leq 1$, Eve's information is restricted. When $p=1$, $I_{BE}= 2-\sqrt{2} \simeq 0.586$ and $I_{AB}$ is equal to 1, giving maximal $K=0.414$. The region where we have non-zero $K$ is $0.9038 \leq p \leq 1$. The key generation rate we obtained is the same as optimal rate found in Eq. (8) in Ref. \cite{Aci06}.
\section{Conclusion}
We outlined a straightforward approach for obtaining a secret key rate using only no-signaling constraints and linear programming. Assuming an individual attack, we considered all possible joint probabilities. We initially examined the case where Eve has binary outcomes. We imposed constraints due to the no-signaling principle and given measurement outcomes. Within the remaining space of joint probabilities, by using linear programming, we optimized the guessing probability between Bob and Eve. We then presented an inequality that relates the guessing probability to the mutual information between Bob and a general Eve who is not binary-restricted. Using the bound and the Csisz\'ar-K\"orner formula \cite{Csi78}, we lower bounded the final key generation rate. The optimal value of the key generation rate, obtained in the noiseless case $p=1$, exactly matches the result from Ref. \cite{Aci06}. However, our approach does not require any specific knowledge of the no-signaling polytopes, instead relying on linear programming techniques to optimize the relevant quantities. Thus, our approach holds promise for application to other protocols, where the structure of the no-signaling polytopes cannot be determined analytically.

\section*{Acknowledgement}
This study was supported by Basic Science Research Program through the National Research Foundation of Korea (NRF) funded by the Ministry of Education, Science and Technology (2010-0007208), and by National Research Foundation and Ministry of Education, Singapore. NK acknowledges the Ontario Graduate Scholarship program for support.


\begin{references}
\bibitem{AMP} With respect to physical implementation, the protocol is almost the same as the Ekert protocol \cite{Eke91}. However, because security is analyzed with a different, though related, point of view, we give a new name.
\end{references}
\end{document}